\documentclass[authoryear,11pt]{elsarticle}
\usepackage{amsmath,amsfonts,amssymb,amsthm,hyperref}
\usepackage{epsfig}
\usepackage{algorithm}
\usepackage{algpseudocode}
\usepackage{color}

\journal{Computational Statistics \& Data Analysis}

\def\mR{\mathbb{R}}

\def\sign{\mbox{sign}}

\def\diag{\mbox{diag}}

\def\wvec{\mbox{vec}}

\def\proc{\mbox{proc}}
\def\soft{\mbox{soft}}

\newcommand{\bSig}{\mbox{\boldmath $\Sigma$}}

\newcommand{\bI}{\mathbf I} 
 
\newcommand{\bzero}{\mathbf 0} 

\newcommand{\A}{\mathbf A}
\newcommand{\B}{\mathbf B}

\newcommand{\X}{\mathbf X}
\newcommand{\Y}{\mathbf Y}

\newcommand{\w}{\mathbf w}
\newcommand{\x}{\mathbf x}
\newcommand{\y}{\mathbf y}
\newcommand{\z}{\mathbf z}

\newcommand{\ba}{\mathbf a}
\newcommand{\bb}{\mathbf b}
\newcommand{\bu}{\mathbf u}
\newcommand{\bv}{\mathbf v}

\newcommand{\aU}{\mathbf U}
\newcommand{\aD}{\mathbf D}

\newcommand{\bU}{\mathbf U}
\newcommand{\bV}{\mathbf V}

\newcommand{\cX}{\mathbb {X}}

\newcommand{\trans}{^\top}

\DeclareMathOperator*{\argmin}{arg\,min} 
\def\defby{\stackrel{\mbox{\textrm{\tiny def}}}{=}}

\newtheorem{prop}{Proposition}[section]

\usepackage{numcompress}
\begin{document}
	
	\begin{frontmatter}
		\title{HiQR: An efficient algorithm for high-dimensional quadratic regression with penalties}
\author[a1]{Cheng Wang\corref{mycorrespondingauthor}}
\ead{chengwang@sjtu.edu.cn}
\cortext[mycorrespondingauthor]{Corresponding author}
				
\author[a1]{Haozhe Chen}

\author[a2]{Binyan Jiang}


\address[a1]{School of Mathematical Sciences, MOE-LSC,\\ Shanghai Jiao Tong University,  Shanghai 200240, China.}

\address[a2]{Department of Applied Mathematics, \\The Hong Kong Polytechnic University,
	Hung Hom, Kowloon, Hong Kong.}


		\begin{abstract}  
This paper investigates the efficient solution of penalized quadratic regressions in high-dimensional settings. A novel and efficient algorithm for ridge-penalized quadratic regression is proposed, leveraging the matrix structures of the regression with interactions. Additionally, an alternating direction method of multipliers (ADMM) framework is developed for penalized quadratic regression with general penalties, including both single and hybrid penalty functions. The approach simplifies the calculations to basic matrix-based operations, making it appealing in terms of both memory storage and computational complexity for solving penalized quadratic regressions in high-dimensional settings.
\end{abstract}

		\begin{keyword} ADMM \sep LASSO 
  \sep quadratic regression \sep ridge regression
		\end{keyword}
		
	\end{frontmatter}
\section{Introduction}

Quadratic regression, which extends linear regression by accounting for interactions between covariates, has found widespread applications across various disciplines.  However, as the complexity of the interactions increases quadratically with the number of variables, parameter estimation becomes increasingly challenging for problems with large or even moderate dimensionality.  A surge of methodologies have been developed in the past decade to tackle the high-dimensionality challenge under different structural assumptions; see for example \cite{bien2013lasso, hao2014interaction, hao2017note, hao2016model, tang2020high, wang2021penalized, lu2021} and  \cite{yu2019reluctant}, among others. 

Given the observations $(\x_i,y_i) \in \mR^{p}\times \mR,~i=1,\ldots,n$, we consider a general penalized quadratic regression model expressed as
\begin{align}\label{pqr}
	\argmin_{\B=\B^\top,  \B\in \mR^{p \times p}} \frac{1}{2n} \sum_{i=1}^n (y_i-\x_i\trans \B \x_i)^2+f(\B),
\end{align}
where $\B=(B_{jk})_{p \times p}$ denotes a symmetric matrix of parameters, and $f(\cdot)$ is a convex penalty function.  Typically, the first element of $\x_i$ is a constant 1, allowing for the capture of the intercept, linear effect, and interaction effect through $B_{1,1,}$, $\{B_{1,i}, i=2,\ldots, p\}$, and $\{B_{i,j}, 2\le i\le j \le p\}$, respectively.

Without the penalty $f(\B)$, the mean squared error is: 
 \begin{align*}
\frac{1}{2n} \sum_{i=1}^n \left(y_i-B_{11}-\sum_{j=2}^p (B_{1j}+B_{j1})x_{ij}-\sum_{j,k=2}^p B_{jk} x_{ij}x_{ik} \right)^2.
\end{align*}
The penalty term $f(\B)$ is introduced to impose different structures on the parameter matrix $\B$ depending on the application scenario.  For instance, in gene-gene interaction detection where the number of genes is typically large and the interactions related to the response are sparse, the $\ell_1$ penalty $f(\B)=\lambda \|\B\|_1$ is often used to induce sparsity in $\B$.  The resulting model is called the all-pairs LASSO by \cite{bien2015convex}. 
 In addition to sparsity, researchers have also considered heredity, where the existence of the interaction effect $B_{j,k}$ depends on the existence of its parental linear effects $B_{1,j}, B_{1, k}$. Specifically, we have: 
\begin{align*}
\mbox{strong heredity:}~B_{j,k} \neq & 0 \Rightarrow B_{1j} \neq 0 ~\mbox{and}~ B_{1k} \neq 0,\\
\mbox{weak heredity:}~B_{j,k} \neq & 0 \Rightarrow B_{1j} \neq 0 ~\mbox{or}~ B_{1k} \neq 0.
\end{align*}
Several penalty functions are proposed in the literature to enforce these heredity structures, including those proposed by  \cite{yuan2009structured}, \cite{radchenko2010variable}, \cite{choi2010variable}, \cite{bien2013lasso}, \cite{lim2015learning}, \cite{haris2016convex}, and  \cite{she2017group}, among others.
 In addition to sparsity and heredity,  we can also introduce the  nuclear norm penalty  to impose a low rank structure in $\B$, and hybrid penalties to impose more than one structure.  Further details will  be provided in Section 3. 

A naive approach to solving the penalized quadratic regression model \eqref{pqr} is to use vectorization. We define
\begin{align*}
	\z_i=\x_i \otimes \x_i \in \mR^{p^2 \times 1},
\end{align*}
where \(\otimes\) denotes the Kronecker product, and write 
\begin{align*}
	\bb=\wvec(\B) \in \mR^{p^2 \times 1},
\end{align*}
where $\wvec(\cdot)$ denotes the vectorization of a matrix. We can then obtain  the following equivalent form of \eqref{pqr}:
\begin{align*}
\argmin_{ \bb } \frac{1}{2n} \sum_{i=1}^n (y_i-\z_i\trans \bb)^2+f(\bb).
\end{align*}
Therefore, the penalized quadratic regression problem \eqref{pqr} can be reformulated as a penalized linear model with $O(p^2)$ features. From a theoretical perspective, we can use this formulation together with the classical theory for high-dimensional regularized $M$-estimators \citep[Chapter 9]{wainwright2019high}. Detailed theoretical analyses of the consistency of the penalized quadratic regression model can be found in \cite{zhao2016analysis} and the references therein.
However, from a computational perspective, many algorithms do not scale well with a large $p$, since the number of parameters scales quadratically with the dimension $p$. Moreover, storing the design matrix and computer memory can also be expensive when vectorization is applied to the interaction variables. For example, computing an all-pairs LASSO with $n=1000$ and $p=1000$ on a personal computer can cause the well-known algorithm \textit{glmnet} \citep{glmnet} to break down due to out-of-memory errors. Specifically, the feature matrix of order $10^3 \times 10^6$ has a memory size of about 8GB.

To address the computational challenges associated with high-dimensional penalized quadratic regression, several two-stage methods have been proposed in the literature \citep[e.g.,]{hao2014interaction, fan2015innovated, kong2017interaction, hao2016model, yu2019reluctant}. These methods are computationally efficient and have been proven to be consistent under some structural assumptions, which can reduce the computational complexity via a feature selection procedure in the first stage. In this paper, we do not assume any of these structures, and our main goal is to develop efficient algorithms for solving the general penalized quadratic regression model \eqref{pqr} directly.
Intuitively, penalized quadratic regression is different from a common linear regression with $O(p^2)$ features because the data has a specific structure for interactions. In this work, we leverage this structure in the algorithm and design an efficient framework for the general penalized quadratic regression problem. In previous works, \cite{tang2020high}  and \cite{wang2021penalized}  also developed efficient formulas for the matrix parameter under a factor model.  However, their procedures greatly rely on the distributional assumptions and cannot be extended to general cases.
   In contrast, our approach does not require any distributional assumptions and can be applied to a wide range of high-dimensional data.

In this work, we study the original optimization problem \eqref{pqr} and design the algorithm from the viewpoint of matrix forms. To the best of our knowledge, this is the first algorithm for penalized quadratic regression that does not use vectorization and avoids any matrix operation of the $n \times p^2$ feature matrix. Our contributions are summarized as follows:

\begin{enumerate}
\item For ridge regression, we obtain an efficient solution for quadratic regression with a computational complexity of $O(np^2+n^3)$.  
\item 
To solve the general penalized quadratic regression problem for single non-smooth penalty and hybrid penalty functions, we propose an alternating direction method of multipliers (ADMM) algorithm. The algorithm is fully formulated with matrix forms, using only $p \times p$, $n \times p$, or $n \times n$ matrices, and has explicit formulas for the solutions in each iteration. 

\item    
We have developed an R package for penalized quadratic regression. Compared to other existing solvers/packages, our algorithm is much more robust since we do not impose any structural assumptions such as heredity or distributional conditions. Our algorithm is appealing in both memory storage and computational cost, and can handle datasets with very high dimensions. This makes our package a useful tool for researchers and practitioners who need to analyze high-dimensional data using penalized quadratic regression.
\end{enumerate}

The rest of the paper is organized as follows. In Section 2, we start with ridge-penalized quadratic regression and derive an efficient closed-form formula for the solution. In Section 3, we design an efficient ADMM algorithm for both single non-smooth penalty and hybrid penalty functions. We conduct simulations in Section 4 to illustrate the proposed algorithm and conclude the work in Section 5 with discussions. The developed R package ``HiQR" and all the codes for simulations are available on GitHub at \url{https://github.com/cescwang85/HiQR}.

\section{Ridge regression}
 
To facilitate the discussion, we introduce some notations first. For a real $p \times q$ matrix $\A = (A_{k,l})_{p \times q}$, we define: 
\begin{align*}
 	\|\A\|_{\infty} \defby \max_{1\le k\le p,1\le l\le q}|A_{k,l}|,~  \|\A\|_{1} \defby\sum_{k=1}^p \sum_{l=1}^q|A_{k,l}|,~\|\A\|^2_{2} \defby\sum_{k=1}^p \sum_{l=1}^q|A_{k,l}|^2.
 \end{align*}
Denoting the singular values of $\A$ as $\sigma_1 \geq \cdots \sigma_p \geq 0$, the nuclear norm of $\A$ is defined as
\begin{align*}
	\|\A\|_*=\sum_{i=1}^p \sigma_i.
\end{align*}

We first consider the ridge regression for the quadratic regression, i.e., 
\begin{align}\label{RidgeQR}
\mbox{Ridge QR:}~~\argmin_{\B=\B^\top, \B \in \mR^{p \times p}} \frac{1}{2n} \sum_{i=1}^n (y_i-\x_i\trans \B \x_i)^2+\frac{\lambda }{2} \|\B\|_2^2.
\end{align}
where $\lambda>0$ is a tuning parameter.
Since the object function is convex in $\B$, the solution can be obtained by solving the following equation:
\begin{align}\label{ridge_eq}
\frac{1}{n} \sum_{i=1}^n (\x_i\trans \B \x_i-y_i) \x_i \x_i \trans +\lambda  \B=\bzero_{p \times p}.
\end{align}
Denote
$ \aD=\frac{1}{n} \sum_{i=1}^n y_i \x_i \x_i \trans$. 
Equation \eqref{ridge_eq} can be equivalently written as: 
\begin{align*}
	\frac{1}{n} \sum_{i=1}^n \x_i \x_i\trans \B \x_i \x_i\trans+\lambda  \B=\aD. 
\end{align*}
By applying vectorization to the above equation, we have:
\begin{align*}
	\frac{1}{n} \sum_{i=1}^n \left\{ (\x_i \x_i\trans) \otimes (\x_i \x_i\trans) \right\} \wvec(\B)+\lambda \cdot \wvec(\B)=\wvec(\aD),
\end{align*}
and then the solution can be seen as:
\begin{align} \label{sol}
	\wvec(\B)= \left\{\cX \cX \trans+\lambda  \bI_{p^2} \right\}^{-1} \wvec(\aD)=\left\{\cX \cX \trans+\lambda   \bI_{p^2} \right\}^{-1} \cX  \Y,
\end{align}
where 
\begin{align*}
	\cX	=\frac{1}{\sqrt{n}} (\x_1 \otimes \x_1,\cdots,\x_n \otimes \x_n).
\end{align*}
Note that $\cX \cX \trans$ is a $p^2 \times p^2$ matrix, which can lead to a high computational complexity of $O(p^6)$ for direct calculation of its inverse. Moreover, storing such a large matrix when $p$ is large is also impractical.  Therefore, the naive algorithm that computes \eqref{sol} directly is usually not applicable for high-dimensional quadratic regression.

Note that the rank of $\cX$ is $\min\{n,p^2\}$, which can be much smaller than $p^2$ when $n \ll p$. To exploit the low-rank structure of $\cX$, we can use the Woodbury matrix identity, which allows us to compute $(\cX \cX \trans + \lambda  \bI_{p^2})^{-1}$ more efficiently. Specifically, by applying the Woodbury identity,
we have:
\begin{align} \label{wood}
	(\cX \cX \trans+\lambda   \bI_{p^2})^{-1}=\lambda ^{-1}\bI_{p^2}-\lambda ^{-1} \cX (\lambda \bI_n+\cX\trans \cX)^{-1} \cX \trans.
\end{align}
The computational complexity is now been reduced to $O(n^2p^2+n^3)$, where the $n^2p^2$ term is due to matrix multiplication and $n^3$ is the complexity of matrix inverse. The Woodbury identity has been widely used in many other algorithms, and it is sometimes referred to as the ``shortcut-trick" for high-dimensional data (\citep[section 4.2.4]{boyd2011distributed}; \citealt{friedman2001elements}).

 Another efficient technique to further reduce the computational cost is the  implementation of the singular value decomposition(SVD) to $\cX$ \citep{haris2016convex}.  Specifically, let $\cX = \bU \Lambda \bV \trans$ be the thin SVD of $\cX$. Together with \eqref{wood}, the solution \eqref{sol} can be expressed as:
\begin{align} \label{svd}
	(\cX \cX \trans+\lambda   \bI_{p^2})^{-1}\cX  \Y=\bU (\Lambda^2+\lambda \bI)^{-1} \Lambda \bV \trans \Y.
\end{align}
Here, the complexity of SVD is $O(n^2p^2)$, which can significantly reduce the computational complexity compared to the naive algorithm that computes \eqref{sol} directly. However, for some large-scale problems, the reduction in computational complexity may still be insignificant.

In what follows, we will further exploit the special structure of the parameter matrix in quadratic regression and reduce the computational complexity to $O(np^2)$. Note that from \eqref{wood} and the first equation of \eqref{sol}, we have:
\[
\wvec(\B)= \left\{ 
\lambda ^{-1}\bI_{p^2}-\lambda ^{-1} \cX (\lambda \bI_n+\cX\trans \cX)^{-1} \cX \trans
\right\}  \wvec(\aD).
\]
Firstly,  note that
\begin{align}\label{p1}
	\cX \trans \cX=&\begin{pmatrix}
		\frac{1}{\sqrt{n}} (\x_1 \otimes \x_1)\trans\\
		\vdots\\
		\frac{1}{\sqrt{n}} (\x_n \otimes \x_n)\trans \nonumber\\
	\end{pmatrix}\left(\frac{1}{\sqrt{n}}  \x_1 \otimes \x_1,\cdots,\frac{1}{\sqrt{n}}  \x_n \otimes \x_n\right)\\
	=&\frac{1}{n} \left( (\x_i \trans \x_j)^2  \right)_{n \times n}=n^{-1}  (\X \X \trans )\circ (\X  \X \trans ),
\end{align}
where $\circ$ is the Hadamard  product and the complexity of the last equation is of order $O(n^2p)$. Secondly, note that
\begin{align}\label{p2}
	\cX \trans \wvec(\aD)=&
	\begin{pmatrix}
		\frac{1}{\sqrt{n}} (\x_1 \otimes \x_1)\trans\\
		\vdots\\
		\frac{1}{\sqrt{n}} (\x_n \otimes \x_n)\trans\\
	\end{pmatrix}  \wvec(\aD)
	=
	\begin{pmatrix}
		\frac{1}{\sqrt{n}} \x_1\trans \aD \x_1\\
		\vdots
		\\
		\frac{1}{\sqrt{n}} \x_n\trans \aD \x_n\\
	\end{pmatrix}  \nonumber\\
	=&\frac{1}{\sqrt{n}} \cdot \diag\left(  \X  \aD \X \trans \right),
\end{align}
where in the last equation the complexity is also reduced to $O(np^2)$. Lastly, denoting
\begin{align*}
	\w=\frac{1}{\sqrt{n}}(\lambda \bI_n+\cX\trans \cX)^{-1} \cX \trans  \wvec(\aD) \in \mR^{n},
\end{align*}
we have: 
\begin{align}\label{p3}
	\cX (\lambda \bI_n+\cX\trans \cX)^{-1} \cX \trans  \wvec(\aD)=&\sum_{k=1}^n w_k \x_k \otimes \x_k \nonumber \\
	=&\wvec\left( \sum_{i=1}^n w_k \x_k \x_k \trans \right)=\wvec \left( \X \trans \diag(\w) \X  \right),
\end{align}
where the complexity of the last equation is also  $O(np^2)$.

By combining equations \eqref{p1}-\eqref{p3}, we can obtain a computationally efficient form for the explicit solution of the ridge-penalized quadratic regression \eqref{RidgeQR}. We summarize the results in the following proposition.
\begin{prop} \label{prop1}
For a given tuning parameter $\lambda>0$, the solution of the ridge-penalized quadratic regression problem \eqref{RidgeQR} is given as:
	\begin{align}\label{eff_sol}
		\widehat{\B}=\lambda^{-1} \aD-\lambda^{-1} \X \trans \diag\{\w\} \X,
	\end{align}
	where 
	\begin{align*}
		\X=&(\x_1,\ldots,\x_n),\quad \aD=\frac{1}{n} \sum_{i=1}^n y_i \x_i \x_i \trans =\frac{1}{n}\X \trans \diag(y_1,\cdots,y_n) \X,\\
			\w=&\left\{ \lambda \bI_n+n^{-1}  (\X \X \trans )\circ (\X  \X \trans ) \right\}^{-1} \diag\left(\frac{1}{n}  \X \aD \X \trans\right).
	\end{align*}
\end{prop}
The computational complexity for calculating the close-form solution \eqref{eff_sol} is $O(np^2+n^3)$, which is  much more efficient than the forms given in \eqref{sol} and \eqref{svd} under the high-dimensional setting where $n\ll p$.  In addition, the memory cost of the solution is also lower because it only requires components in the form of either an $n \times n$ matrix, an $n \times p$ matrix, or a $p \times p$ matrix. In next section, we will further extend our results obtained in this section to solve quadratic regression with other non-smooth penalties.

\section{Non-smooth penalty and beyond}

In this section, we consider the case where the penalty $f(\cdot)$ in the penalized quadratic regression \eqref{pqr} is possibly non-smooth. For example, we can consider setting $f(\B)=\lambda \|\B\|_1$ as in the all-pairs-LASSO, or $f(\B)=\lambda \|\B\|_*$ as in reduced rank regression. For high-dimensional quadratic regression, it is also attractive to introduce additional penalties to impose different structures simultaneously. For instance, we can combine the $\ell_1$ norm and the nuclear norm to get a sparse and low-rank solution, i.e., $f(\B)=\lambda_1 \|\B\|_1+\lambda_2 \|\B\|_*$. In the literature, several hybrid penalty functions are proposed for quadratic regression, and we summarize these hybrid penalties as follows.

\begin{itemize}
	\item $\ell_1+\ell_2$: 
\begin{align*}
	f(\B)=\lambda_1 \|\B\|_1+\lambda_2 \sum \limits_{k=2}^p \|\B_{\cdot,k}\|_2+\lambda_2 \sum \limits_{k=2}^p \|\B_{k,\cdot}\|_2.
\end{align*}
See \cite{radchenko2010variable} and \cite{lim2015learning} for more details. 
\item $\ell_1+\ell_\infty$:
\begin{align*}
	f(\B)=\lambda_1 \|\B\|_1+\lambda_2 \sum \limits_{k=2}^p \|\B_{\cdot,k}\|_\infty+\lambda_2 \sum \limits_{k=2}^p \|\B_{k,\cdot}\|_\infty.
\end{align*} 
See \cite{haris2016convex}.
\item $\ell_1+\ell_1/\ell_\infty$:
\begin{eqnarray*}
f(\B)&=&\lambda_1 \|\B\|_1+\lambda_2 \sum \limits_{k=2}^p \max\{|\B_{1,k}|, \|\B_{-1,k}\|_1\} \\
&&+\lambda_2 \sum \limits_{k=2}^p \max\{|\B_{k,1}|, \|\B_{k,-1}\|_1\}.
\end{eqnarray*}
See \cite{bien2013lasso} and \cite{haris2016convex}.  
	\item $\ell_1+\ell_*$: 
\begin{align*}
	f(\B)=\lambda_1 \|\B\|_1+\lambda_2 \|\B\|_*.
\end{align*}	
See \cite{lu2021} and  the references therein. 
\end{itemize}  
We remark that all of these penalties are formulated in a symmetric pattern, i.e., $f(\B)=f(\B\trans)$. Thus, the final solution will be a symmetric matrix. 
Utilizing the efficient formulation we obtained in Proposition \ref{prop1}, we now introduce an ADMM algorithm for solving the general penalized quadratic regression problem \eqref{pqr}.

\subsection{ADMM algorithm}
Writing the squared loss function
\begin{align*}
	f_0(\B)=\frac{1}{2n} \sum_{i=1}^n (y_i-\x_i \trans \B \x_i)^2, 
\end{align*}
we study the generic problem 
\begin{align*}
	\min f_0(\B)+f_1(\B)+\cdots+f_N(\B),
\end{align*}
where $f_k(\cdot),k=1,\ldots,N$ are penalty functions. Introducing the local variables $\B_i \in \mR^{p \times p}$, 
the problem can be equivalently rewritten as the following \textit{global consensus problem} \citep[Section 7]{boyd2011distributed}
\begin{align} \label{gen1}
	\min \sum_{i=0}^N f_i(\B_i),~\mbox{subject~to~}~\B_i-\B=\mathbf{0},~i=0,1,\ldots, N.
\end{align} 
The augmented Lagrangian of \eqref{gen1} is
\begin{align*}
L(\B_0,\ldots,\B_N, \B, \bU_0,\ldots,\bU_N)=\sum_{i=0}^N \left\{f_i(\B_i)+ \frac{\rho}{2} \|\B_i-\B+\bU_i\|^2_2   \right\},
\end{align*}
where $\rho>0$ is the step-size parameter. For a given solution $\B_i^{k}, i=0,\ldots, N$ in the $k$th iteration, 
the $(k+1)$th iteration of the ADMM algorithm is given as follow:
\begin{itemize}
 \item Step 1:  	$\B^{k+1}_i=\argmin_{\B_i} \left\{f_i(\B_i)+ \frac{\rho}{2} { \|\B_i-\B^k+\bU^k_i\|^2_2 }   \right\},     i=0,\cdots,N$;  
	\item Step 2:~$\B^{k+1}=\frac{1}{N+1}\sum_{i=0}^N \left\{ \B^{k+1}_i+\bU^k_i \right\}$;
	\item Step 3:~$\bU_i^{k+1}=\bU_i^k+\B_i^{k+1}-\B^{k+1},~i=0,\cdots,N$.
\end{itemize}
If we start with $\sum \bU^1_i=\mathbf{0}$, it can be shown that $\sum \bU^k_i=\mathbf{0}$ for every $k>1$ and so Step 2 will simply be an average operator, i.e.,
\begin{align*}
	\B^{k+1}=\frac{1}{N+1}\sum_{i=0}^N \B^{k+1}_i.
\end{align*}
 As we can see, the computational complexity of the algorithm is usually dominated by the first step. 

In general, for a convex function $f(\cdot)$, the proximal operator \citep{parikh2014proximal} is defined as: 
\begin{align} \label{l2proc}
 \proc_{f, \rho}(\A) 	\defby  \argmin_{\B}	f(\B)+\frac{\rho}{2} \|\B-\A\|_2^2  .
\end{align}
  Thus, given $\B^k$ and the $\bU_i^k$'s,  Step 1 is a proximal operator for the sum of the squared loss function $f_0(\cdot)$ and the penalty functions $f_i(\cdot),~i=1,\ldots,N$. In Proposition \ref{prop1} we have derived an efficient form for the proximal operator of the squared loss $f_0(\cdot)$ at $\A={\bf 0}$. For a general $\A$ in \eqref{l2proc}, the efficient solution can be obtained by  setting $\lambda=\rho/2$ and updating $\aD$ as $n^{-1}\sum_{i=1}^n y_i \x_i \x_i \trans+\A$ in Proposition \ref{prop1}.  In next subsection, we provide the proximal operator for each penalty function.

\subsection{Proximal operator}

For most penalty functions, the proximal projection has an explicit solution, and we summarize these operators in this section. With some abuse of notation, let $\B$ be a parameter matrix with dimension $p\times q$. For the $\ell_1$ norm, writing $\A=(A_{ij})_{p \times q}$, we have
\begin{align*}
	\argmin_{\B}  \lambda \|\B\|_1+\frac{1}{2} \|\B-\A\|_2^2=\left( \sign(A_{ij})( |A_{ij}|-\lambda)_{+} \right)_{p \times q}\defby \soft(\A,\lambda),
\end{align*}
where $x_{+}=\max(0,x)$. For the nuclear norm, denoting the singular value decomposition of $\A$ as  
\begin{align*}
	\A=\sum_{i=1}^{\min(p,q)} \sigma_i \bu_i \bv_i\trans,	
\end{align*}
we have
\begin{align*}
	\argmin_{\B}  \lambda \|\B\|_*+\frac{1}{2} \|\B-\A\|_2^2=\sum_{i=1}^{\min(p,q)} (\sigma_i-\lambda)_{+} \bu_i \bv_i\trans.
\end{align*}

 For other penalties imposed on the columns or the rows of $\B$,  we present the solutions in the form of row vectors for brevity. Without loss of generality, for a convex penalty function $f(\cdot)$ on the row of $\B$, the proximal operator is given as: 
\begin{align*}
	\widehat{\bb}=\argmin_{\bb}	f(\bb)+\frac{1}{2} \|\bb-\ba\|_2^2,~\ba,\bb \in \mR^q,
\end{align*}
we have the following solution.
\begin{itemize}
\item $\ell_2$ norm--Group LASSO \citep{yuan2006model}:
\begin{align*}
	f(\bb)=\lambda \|\bb\|_2 ,~\widehat{\bb}=\left(1-\frac{\lambda}{ \|\ba\|_2} \right)_{+} \cdot \ba.
\end{align*}
\item  $\ell_\infty$ norm penalty \citep{duchi2009efficient}:
\begin{align*}
	f(\bb)=\lambda \|\bb\|_\infty.
\end{align*}
When $\lambda \geq \|\ba\|_1$, we have $\hat{\bb}=\bzero$. Otherwise, the solution is 
\begin{align*}
	\widehat{\bb}=\ba-\soft(\ba,\lambda_1),
\end{align*}
where $\lambda_1 \geq 0$ satisfies the equation
\begin{align*}
	\sum_{i=1}^q (|a_i|-\lambda_1) I(|a_i|>\lambda_1)=\lambda.
\end{align*}
The details of the derivation can be found in Section 5.4 of \cite{duchi2009efficient}.
\item Hybrid $\ell_1/\ell_\infty$ norm penalty  \citep{haris2016convex}:
\begin{align*}
f(\bb)=\lambda \max \left(|b_1|, \sum_{i=2}^q |b_i|\right).
\end{align*} 
The solution is 
\begin{align*}
	\widehat{\y}=\left(\soft(a_1,\lambda_1), \soft(\ba_{-1},\lambda-\lambda_1)\right),
\end{align*}
where 
\begin{align*}
	\lambda_1=\argmin_{t \in [0, \lambda]} \|\soft(a_1,t)\|_2^2+\| \soft(\ba_{-1},\lambda-t)\|_2^2.
\end{align*}
In particular, when $\lambda \geq |a_1|+\|\ba_{-1}\|_\infty$, $\widehat{\bb}=\bzero$. Further details can be found in \cite{haris2016convex}.
\end{itemize}
In the above, we have summarized some commonly used penalties in quadratic regression and their explicit proximal operators. For other non-smooth penalty, the propose algorithm is still applicable and we only need to update the algorithm with the corresponding proximal operators. We remark for the penalties imposed on the column vectors, the  proximal operators can be  obtained similarly.  

With these explicit proximal operators, we can get the unified algorithm as follows.
 \begin{algorithm}[H]\small 
	\caption{HiQR: High dimensional Quadratic Regression.}
	\begin{algorithmic}[1]
		\item[Initialization:]
		\State  Input the observations $(\x_i,y_i),~i=1,\cdots,n$;
		\State  Set the loss function $f_0(\cdot)$ and the penalty functions $f_1(\cdot),\cdots~f_N(\cdot)$;
		\State  Start from $k=0$, $\B^{0}_i=\aU_i^0=\bzero_{p \times p}$.
		\item[Iteration:] 	
		\State  Update $\B^{k+1}_i=\proc_{f_i,\rho}(\B^k-\aU^k_i),~i=0,\cdots,N$.
		\State Update $\B^{k+1}=\frac{1}{N+1}\sum_{i=0}^N \B^{k+1}_i$.
		\State Update $\aU_i^{k+1}=\aU_i^k+\B_i^{k+1}-\B^{k+1}~i=0,\cdots,N$.
		\State 	Repeat steps 4-6 until convergence. 		
		\item[Output:] Return $\B$.
	\end{algorithmic}
\end{algorithm}
Algorithm 1 is simple and efficient owing to the fact that each step of the iteration has a closed form, and we have greatly utilized the matrix structure of the problem to obtain a closed-form solution for the proximal operator of the squared loss for quadratic regression, i.e., in the update of $\B_0$ in Step 4 of Algorithm 1. The algorithm is fully matrix-based, where we update $p \times p$ matrices in each step without any unnecessary matrix operations such as vectorization or Kronecker product. This can greatly reduce the memory and computational burden when handling high-dimensional data.

%

Here we develop the algorithm by following the classical ADMM algorithm \citep{boyd2011distributed} and the convergence results have been explored in the literature. Empirically, the step-size parameter $\rho$ has an impact on the convergence of the algorithm.  Note that there are  $O(p^2)$ parameters and the Hessian matrix $\cX \cX/n \trans$ of the squared loss has  eigenvalues that diverge considerably, we have chosen a relatively large default value, i.e., $\rho=10$, for the step-size parameter in our package.   Alternatively, users can set it to $\rho=\sqrt{p}$.  A more comprehensive way is to  use different step-size parameters for each iteration \cite[e.g., Section 3.4.1]{boyd2011distributed}.

\section{Simulations}
To illustrate the efficiency of the proposed algorithm, we consider a toy example: 
\begin{align*}
Y=2X_1-2X_5+2X_{10}+3X_1X_5-2.5X_5^2+4X_5X_{10}+\epsilon.	
\end{align*}
For all the simulations, we generate $\x_1,\cdots,\x_n$ independently from $N(\bzero,\bSig)$, where $\bSig=(0.5^{|k-l|})_{p \times p}$, and the error term $\epsilon$ from $N(0,1)$.  We fix the sample size $n=500$, and vary the data dimension $p$ from small to large.  The code is implemented on an Apple M1 chip with 8-core CPUs and 8G RAM, and the R version used is 4.3.1 with vecLib BLAS.

\subsection{Ridge regression}
In this part, we compare four algorithms for computing the ridge-penalized quadratic regression, namely, the naive inverse \eqref{sol}, the Woodbury trick \eqref{wood}, the SVD method \eqref{svd}, and the proposed HiQR. We fix $\lambda=10$, and the computation times are recorded in seconds based on 10 replications.

\begin{table}[!htbp] \centering 
	\caption{Average computation time (standard deviation) of different algorithms for ridge regression ($\lambda=10$) over 10 replications. Time is recorded in seconds.  } 
	\label{tab2} 
	\resizebox{1\textwidth}{!}{%
  \begin{tabular}{@{\extracolsep{5pt}} cccccc} 
  \\[-1.8ex]\hline 
  \hline \\[-1.8ex] 
  & p=100 & p=200 & p=400 & p=800 & p=1200 \\ 
  \hline \\[-1.8ex] 
  Naive & 9.814(0.061) &  NA & NA &  NA & NA \\  
Woodbury & 0.131(0.009) & 0.566(0.045) &  2.262(0.173) & 24.772(4.686) & NA \\ 
SVD & 0.534(0.012) & 2.718(0.021) & 13.996(0.095) & 71.996(2.074) & NA \\ 
HiQR & 0.020(0.001) & 0.020(0.001) &  0.024(0.002) &  0.047(0.003) & 0.051(0.004) \\ 
  \hline \\[-1.8ex] 
  \multicolumn{6}{l}{*NA is produced due to out of memory in R.}
  \end{tabular} 
	}
  \end{table} 

From Table \ref{tab2}, we can observe that our HiQR algorithm greatly outperforms other algorithms in terms of computation efficiency.  Additionally, the results are roughly consist with their native computational complexity, e.g., $O(p^6)$, $O(n^2p^2+n^3)$, $O(n^2p^2)$ and $O(np^2+n^3)$. As we can see, the vectorization methods all fail to handle the $p=1200$ case due to   memory shortage, while our method is still efficient, as we only need to handle the storage of $n \times p$ and $p \times p$ matrices.  

\subsection{Single penalty function}
In this part, we investigate the performance of the proposed HiQR for a single penalty, i.e., $f(\B)=\lambda \|\B\|_1$. As a comparison, we also implement the all-pairs LASSO of vectorized features using two state-of-the-art algorithms, e.g., ``glmnet" \citep{friedman2010regularization} and ``ncvreg'' \citep{breheny2011coordinate}.  Table \ref{tab3} reports the computation times of these three algorithms for a solution path with 50 $\lambda$s based on 10 replications. 

\begin{table}[!htbp] \centering  
	\caption{Average computation time (standard deviation) of three packages for obtaining a solution path for all-paris LASSO over 10 replications. The same set of 50 $\lambda$s has been used for the three different packages, and time is recorded in seconds.} 
	  \label{tab3} 
	\resizebox{1\textwidth}{!}{%
	\begin{tabular}{@{\extracolsep{5pt}} cccccccc} 
		\\[-1.8ex]\hline 
		\hline \\[-1.8ex] 
		 & p=200 & p=400 & p=800 & p=1200 & p=1600 & p=2000 & p=2400 \\ 
		\hline \\[-1.8ex] 
		glmnet & 0.65(0.04) & 2.92(0.11) & 21.53(1.89) &  NA &  NA &   NA &   NA \\ 
		ncvreg & 1.38(0.07) & 5.77(0.08) & 34.48(2.96) &  NA &  NA &   NA &   NA \\ 
		HiQR & 1.64(0.54) & 3.68(0.29) & 16.67(1.18) & 45.46(6.13) & 98.78(18.58) & 190.08(46.44) & 298.87(76.66) \\ 
		\hline \\[-1.8ex] 
		\multicolumn{6}{l}{*NA is produced due to out of memory in R.}	
		\end{tabular} 
} 
\end{table}

From Table \ref{tab3}, we can see that both ``glmnet" and ``ncvreg" fail to generate solutions when $p\geq 1200$ due to out-of-memory errors. We note that  ``glmnet" and ``ncvreg" are coordinate descent methods and they use the maximum norm between two iterations to stop the algorithm. Our proposed HiQR is an ADMM method and we use the Frobenius norms of primal and dual errors to stop the iteration. Although the stopping criterion varies for each method, the solutions only differ slightly. In particular, we have checked the stopping condition of our HiQR using the solutions generated from ``glmnet" and ``ncvreg", and found that the 
scales of the stopping condition are comparable to that of the HiQR solution.  Moreover, we remark that both ``glmnet" and ``ncvreg'' are  accelerated by using strong rules; see \cite{tibshirani2012strong} and \cite{lee2015strong} for more details. Strong rules screen out a large number of features to substantially improve computational efficiency. However, as \cite{tibshirani2012strong} has pointed out, the price is that ``the strong rules are not foolproof and can mistakenly discard active predictors, that is, ones that have nonzero coefficients in the solution."   As a comparison, our algorithm can be as efficient as ``glmnet" and ``ncvreg'' without the need for the same type of acceleration.

\subsection{Hybrid penalty functions}
In this part, we report the performance of HiQR for hybrid penalty functions. Specifically, we conduct simulations for the $\ell_1+\ell_2$, $\ell_1+\ell_\infty$, $\ell_1+\ell_1/\ell_\infty$, and $\ell_1+\ell_*$ penalties. The two parameters $\lambda_1$ and $\lambda_2$ are determined by $\lambda$ and $\alpha \in(0,1)$, that is,
\begin{align*}
	\lambda_1=\lambda\cdot\alpha\cdot\lambda_{1,max},~\lambda_2=\lambda\cdot(1-\alpha)\cdot\lambda_{2,max},
\end{align*}  
where $\lambda_{1,max}$ ($\lambda_{2,max}$) is set to be the smallest tuning value corresponding to a zero estimation when $\lambda_2$ $(\lambda_1)$ is set to be 0. We apply HiQR over a $10 \times 10$ grid of $(\alpha,\lambda)$ values, and Table \ref{tab4} presents the average computation times for the whole procedure. As a comparison, we include the ``FAMILY'' method \citep{haris2016convex} which can solve the same problem with $\ell_1+\ell_\infty$  and $\ell_1+\ell_2$ penalties. In the original paper, \cite{haris2016convex} has demonstrated the advantages of these models and here we focus on  the computation time. From Table \ref{tab4}, we can see that the proposed algorithm scales very well to high-dimensional quadratic regression.

\begin{table}[!htbp] \centering  
	\caption{Average computation times (standard deviation) of ``HiQR" and ``FAMILY'' under hybrid penalties with 100 tuning pairs over 10 replications.  Time is recorded in seconds.} 
	  \label{tab4} 
	\resizebox{1\textwidth}{!}{%
	\begin{tabular}{@{\extracolsep{5pt}} ccccc} 
\\[-1.8ex]\hline 
\hline \\[-1.8ex] 
 & Method & p=50 & p=100 & p=200 \\ 
\hline \\[-1.8ex] 
$\ell_1+\ell_\infty$   & HiQR &  2.648(0.216) &    6.592(  0.953) &   23.490(   1.657) \\ 
  & FAMILY & 28.429(1.557) &  259.403(112.138) & NA \\ 
  $\ell_1+\ell_2$ & HiQR &  0.952(0.123) &    1.411(  0.321) &    3.055(   0.231) \\ 
 & FAMILY & 18.820(1.949) & 1399.746(407.768) & NA \\ 
 $\ell_1+\ell_1/\ell_\infty$ & HiQR &  3.065(0.161) &    7.711(  0.824) &   19.878(   1.278) \\ 
 $\ell_1+\ell_*$ & HiQR &  2.285(0.157) &    5.883(  0.510) &   29.322(   2.828) \\ 
\hline \\[-1.8ex] 
\multicolumn{5}{l}{*NA is produced due to FAMILY did not converge.}	
\end{tabular}}
\end{table}

\subsection{Model performance}
Lastly, we evaluate different penalties on different models. In particular, we consider  
\begin{align}
	\mbox{ Model 1:}&~Y=2X_1-2X_5+2X_{10}+3X_1X_5-2.5X_5^2+4X_5X_{10}+\epsilon, \label{m1} \\
	\mbox{ Model 2:}&~Y=-2X_5+3X_1X_5-2.5X_5^2+4X_5X_{10}+\epsilon, \label{m2}	\\
	\mbox{ Model 3:}&~Y=3X_1X_5-2.5X_5^2+4X_5X_{10}+\epsilon, \label{m3}		
	\end{align}
	where the true parameters of $\B[(1,2,6,11),(1,2,6,11)]$ are 
	\begin{align*}
	\begin{pmatrix}
		0&1&-1&1\\
		1&0&1.5&0\\
		-1&1.5&-2.5&2\\
		1&0&2&0
	\end{pmatrix},~\begin{pmatrix}
		0&0&-1&0\\
		0&0&1.5&0\\
		-1&1.5&-2.5&2\\
		0&0&2&0
	\end{pmatrix},\begin{pmatrix}
		0&0&0&0\\
		0&0&1.5&0\\
		0&1.5&-2.5&2\\
		0&0&2&0
	\end{pmatrix},
	\end{align*}
	respectively. 
In particular, Model 1 has a strong hierarchical structure, Model 2 has a weak hierarchical structure, and Model 3 is a model with only interactions.

\begin{figure}[!ht]
	\centerline{
		\begin{tabular}{cccc}				
			\multicolumn{4}{c}{Strong hierarchical model \eqref{m1}}\\
			\psfig{figure=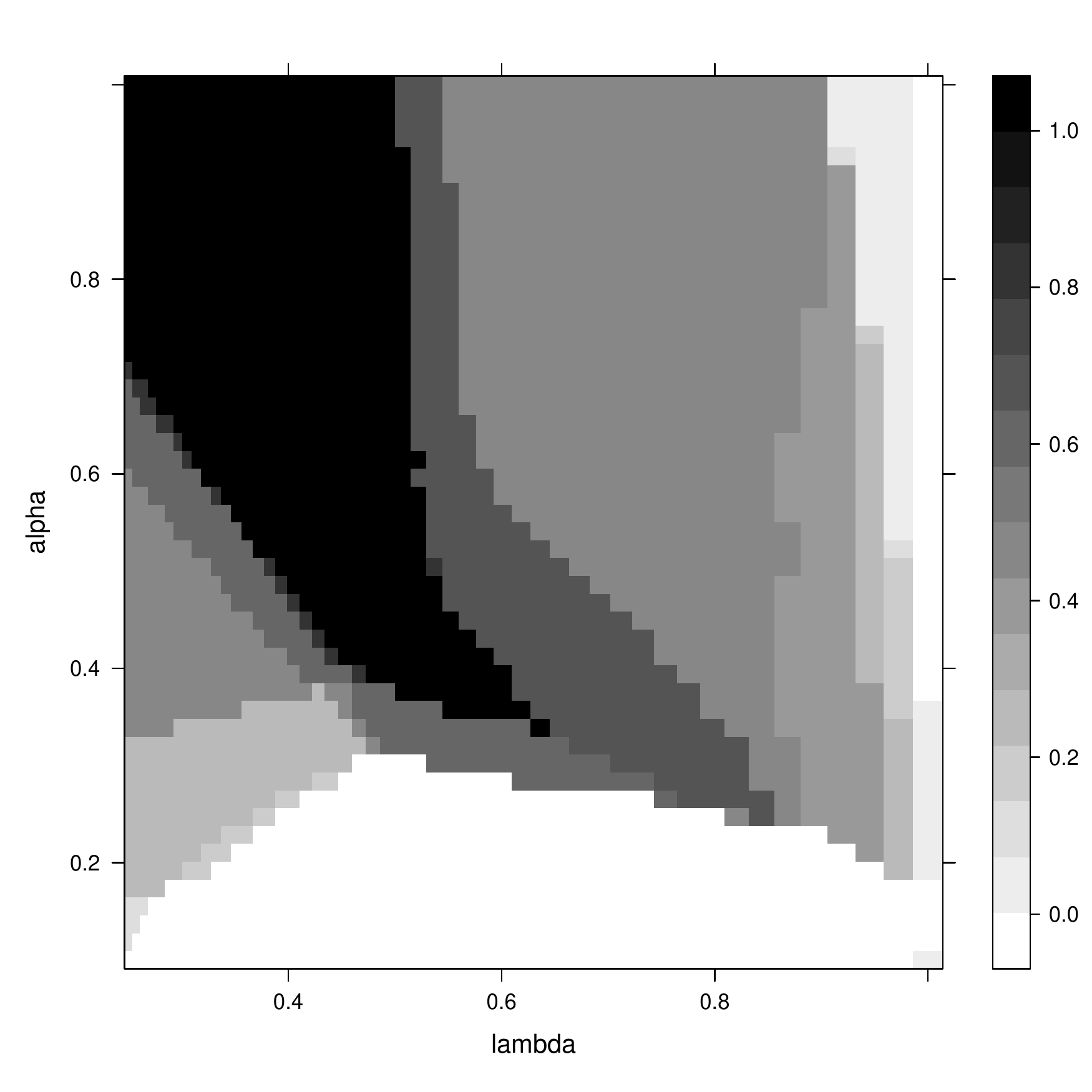,width=1.25 in,height=1.25 in,angle=0} &
			\psfig{figure=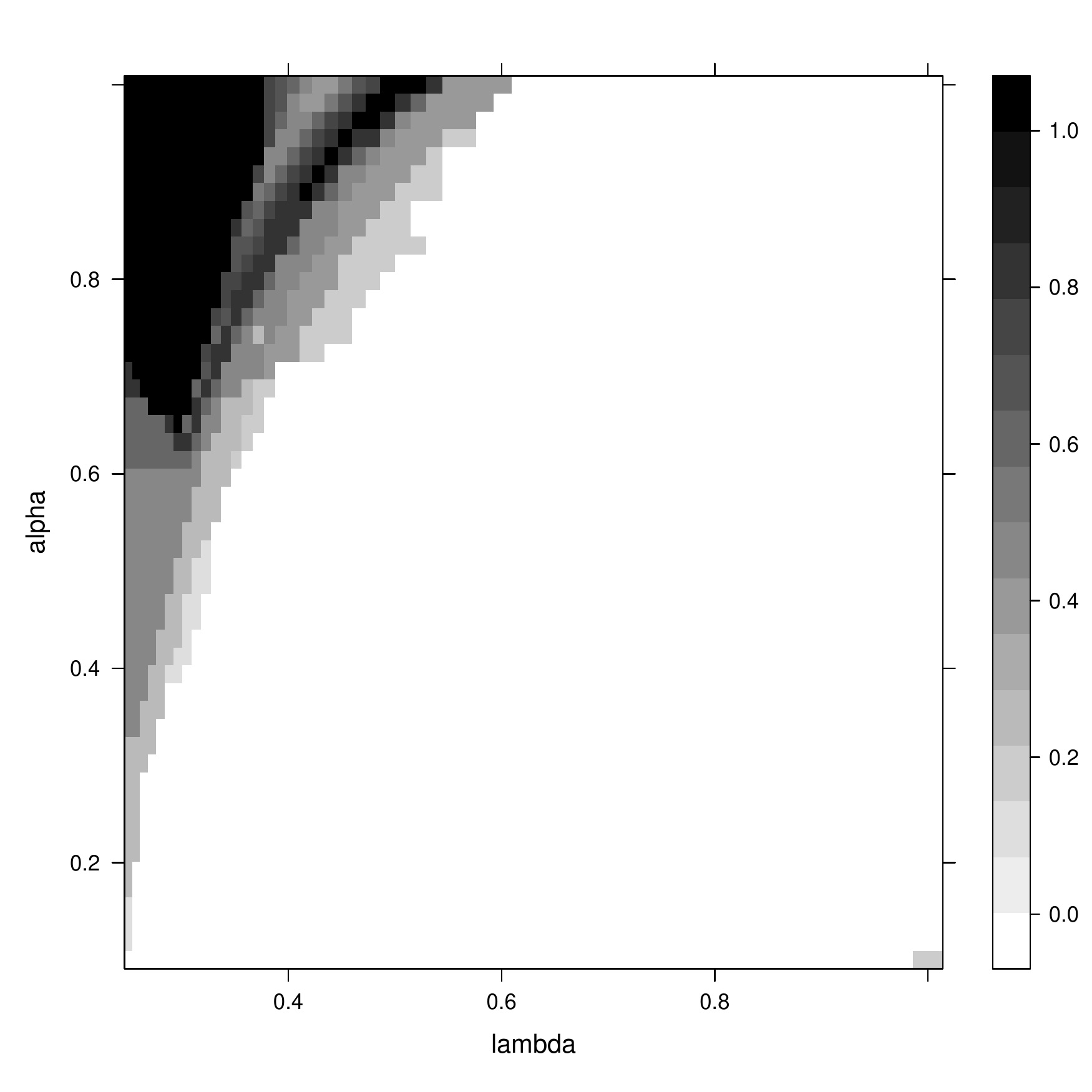,width=1.25 in,height=1.25 in,angle=0} &
			\psfig{figure=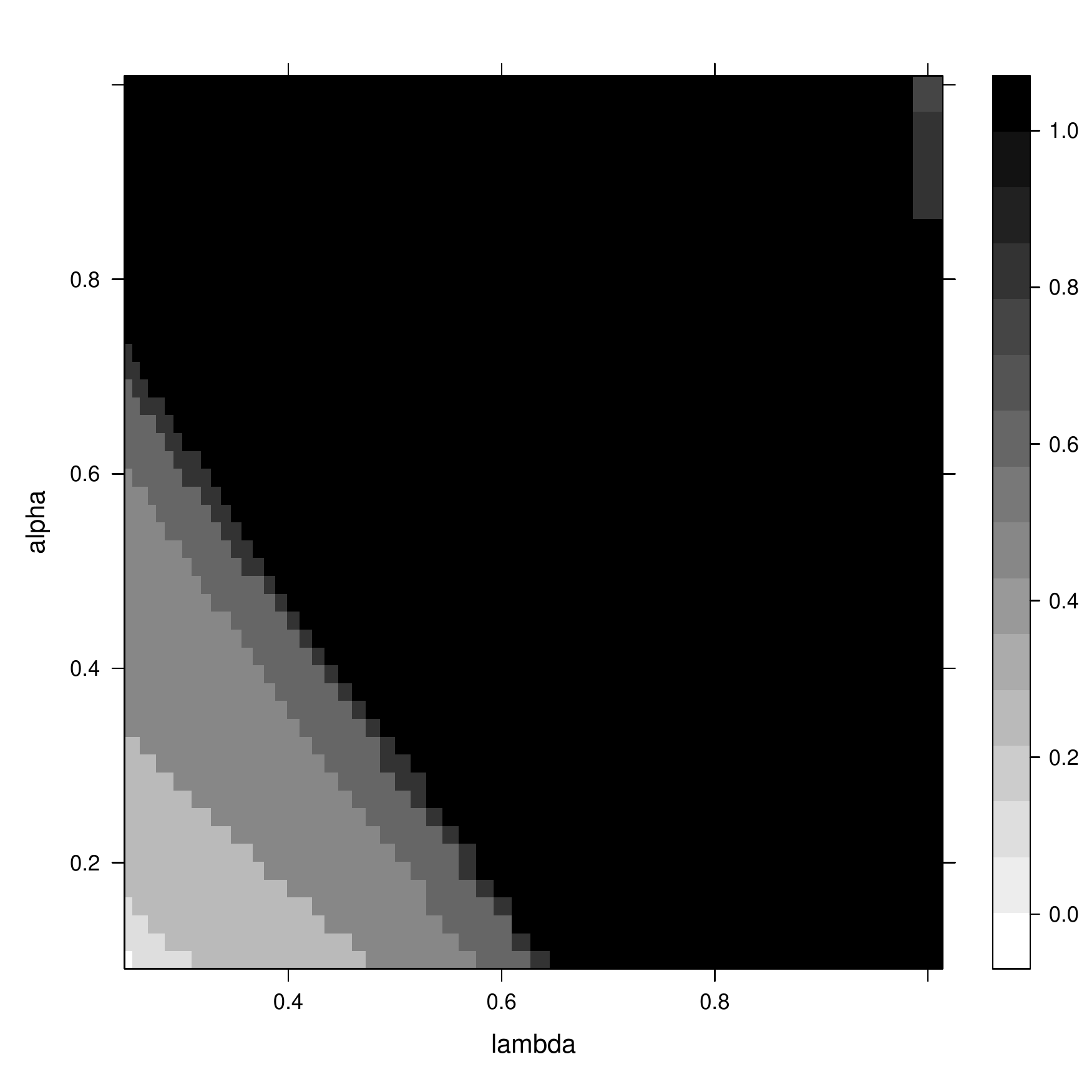,width=1.25 in,height=1.25 in,angle=0} &
			\psfig{figure=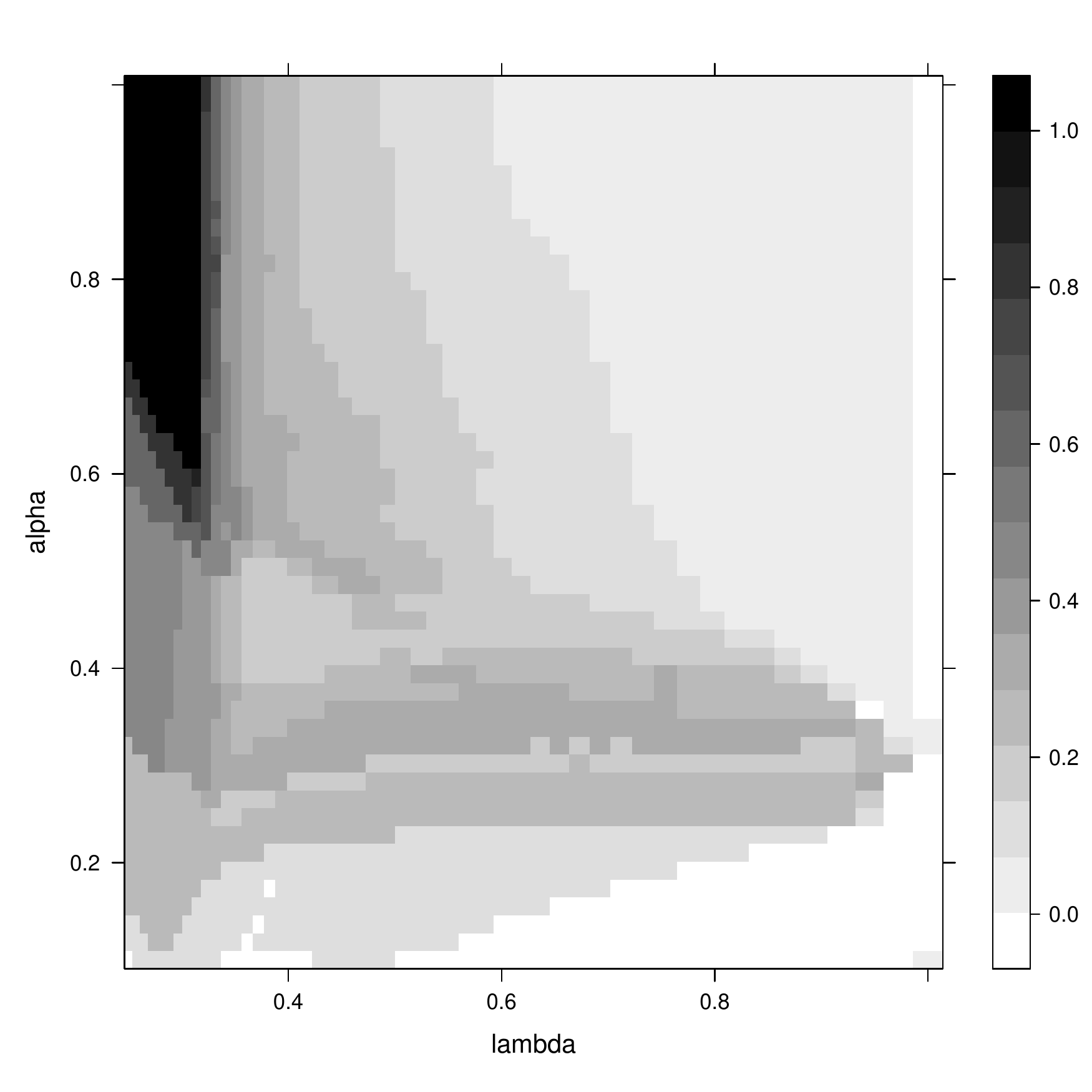,width=1.25 in,height=1.25 in,angle=0} \\
			$\ell_1+\ell_2$ &  $\ell_1+\ell_\infty$ &  $\ell_1+\ell_1/\ell_\infty$&  $\ell_1+\ell_*$\\		
			\hline
			\multicolumn{4}{c}{Weak hierarchical model \eqref{m2}}\\
			\psfig{figure=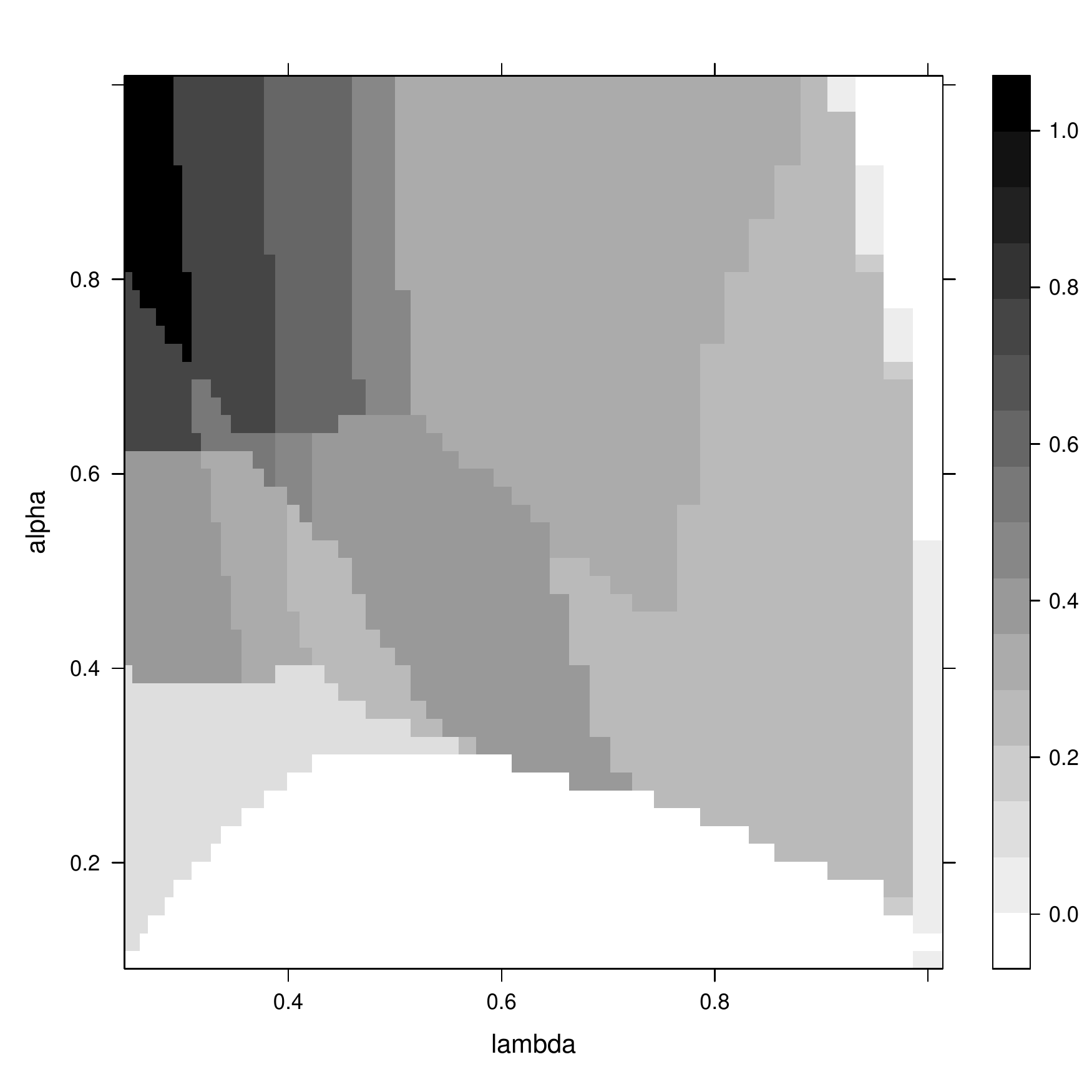,width=1.25 in,height=1.25 in,angle=0} &
			\psfig{figure=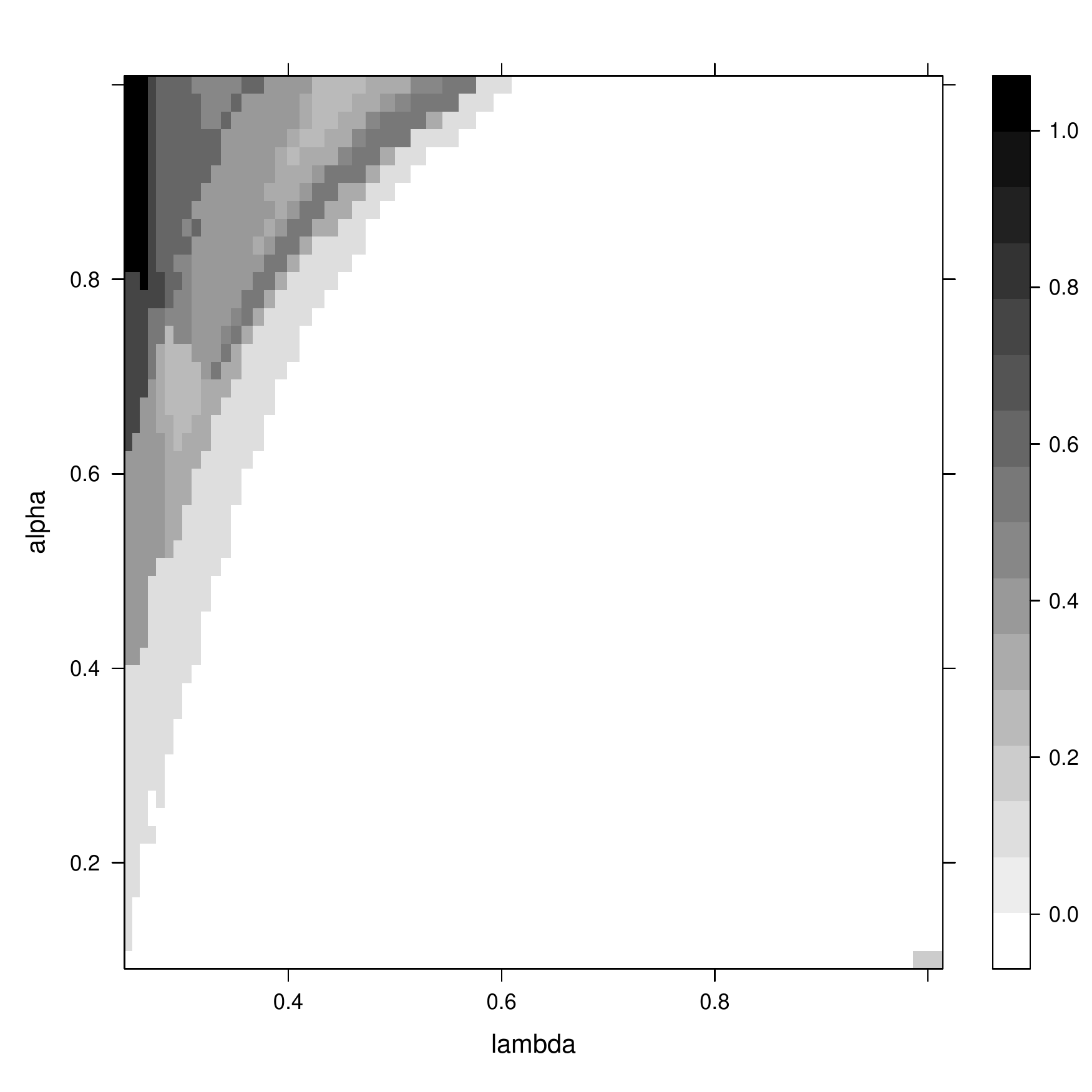,width=1.25 in,height=1.25 in,angle=0} &
			\psfig{figure=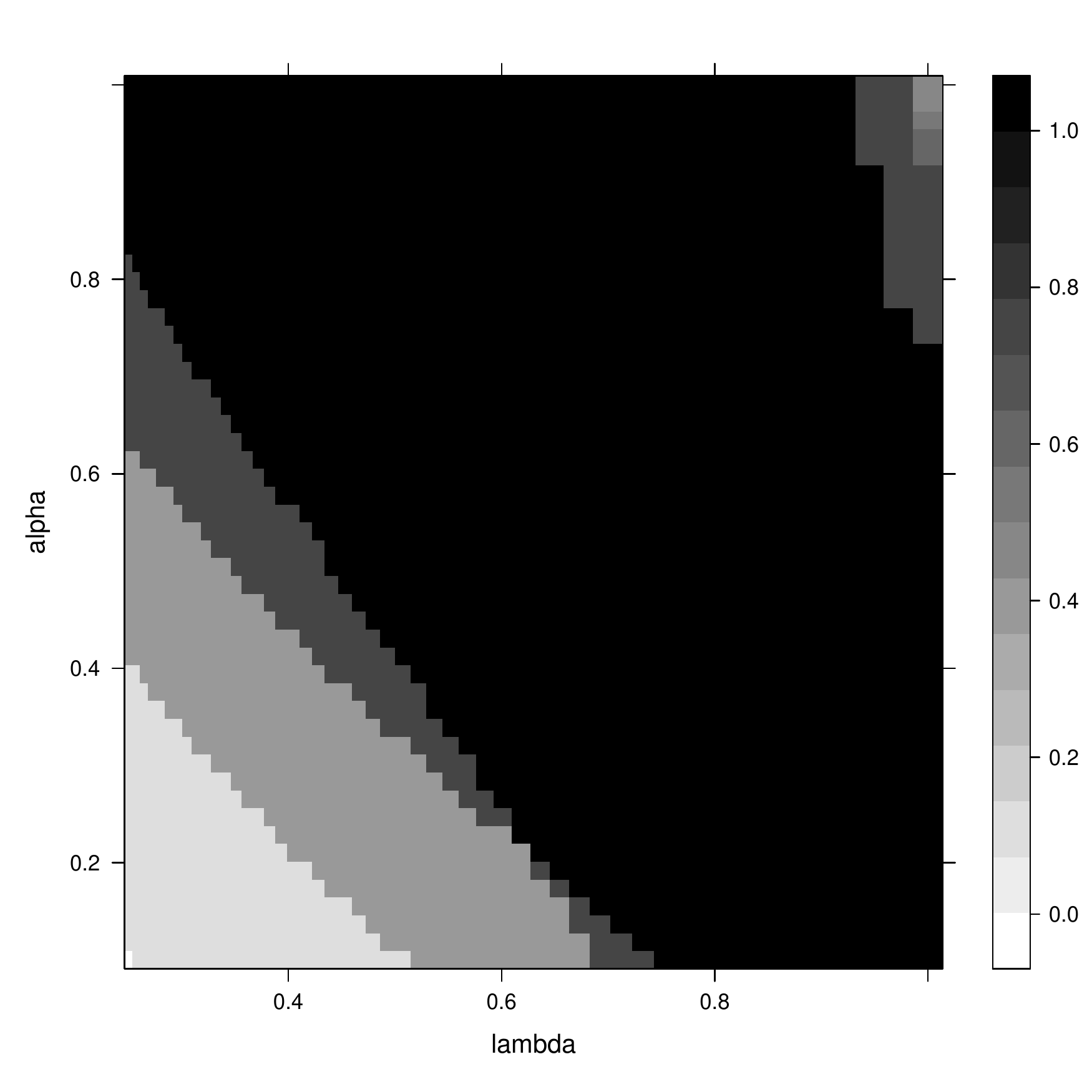,width=1.25 in,height=1.25 in,angle=0} &
			\psfig{figure=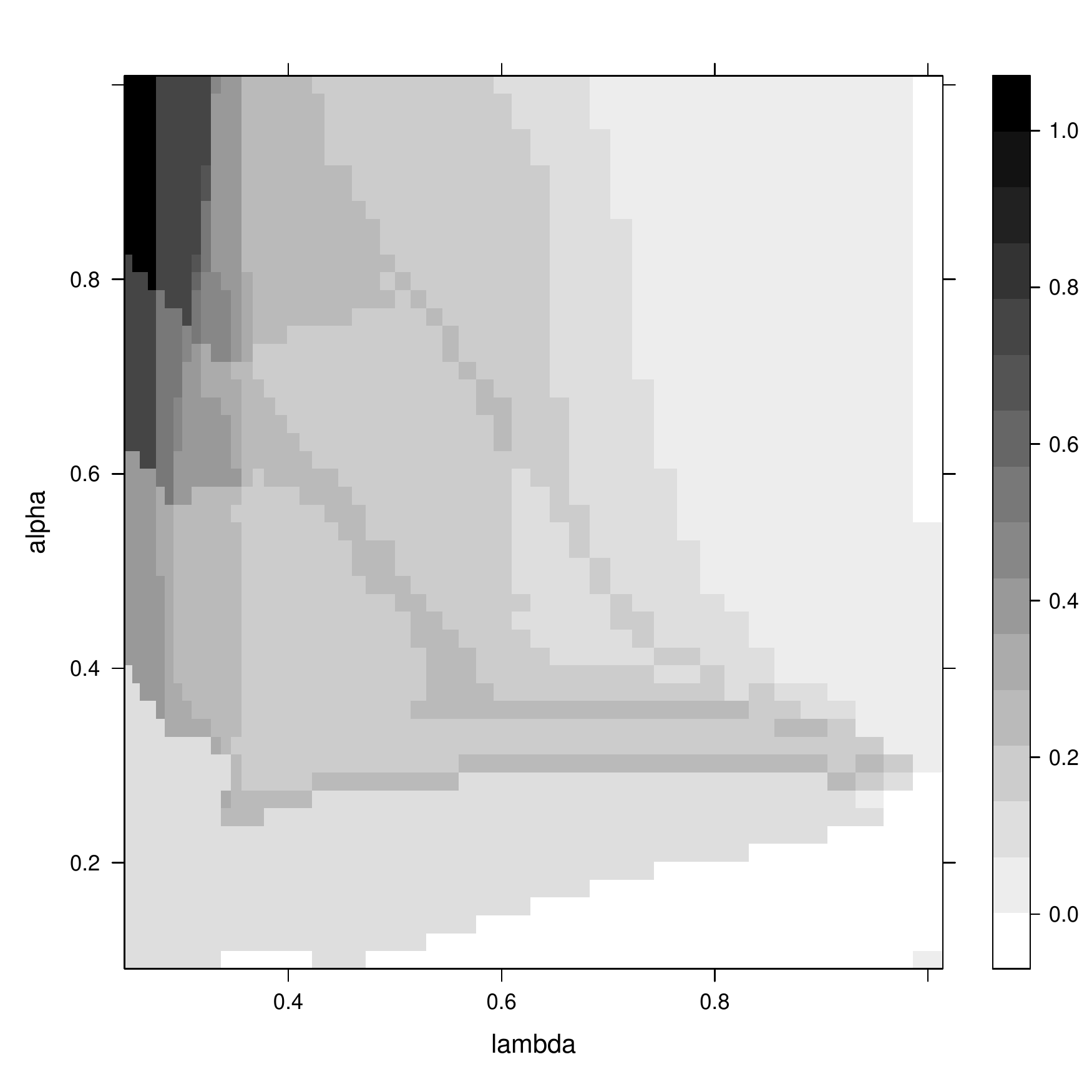,width=1.25 in,height=1.25 in,angle=0} \\
			$\ell_1+\ell_2$ &  $\ell_1+\ell_\infty$ &  $\ell_1+\ell_1/\ell_\infty$&  $\ell_1+\ell_*$\\		
			\hline
			\multicolumn{4}{c}{Pure interaction model \eqref{m3}}\\	
			\psfig{figure=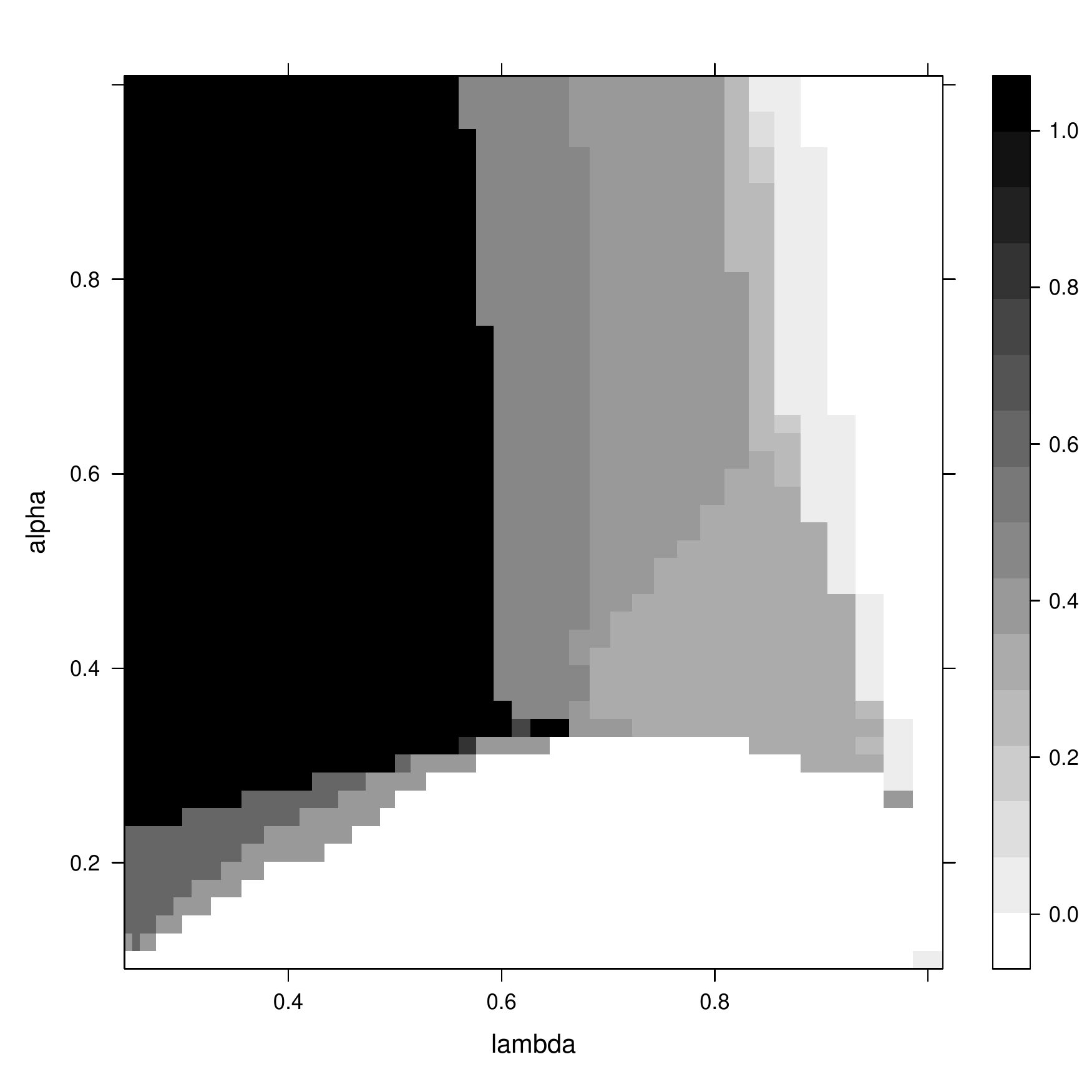,width=1.25 in,height=1.25 in,angle=0} &
			\psfig{figure=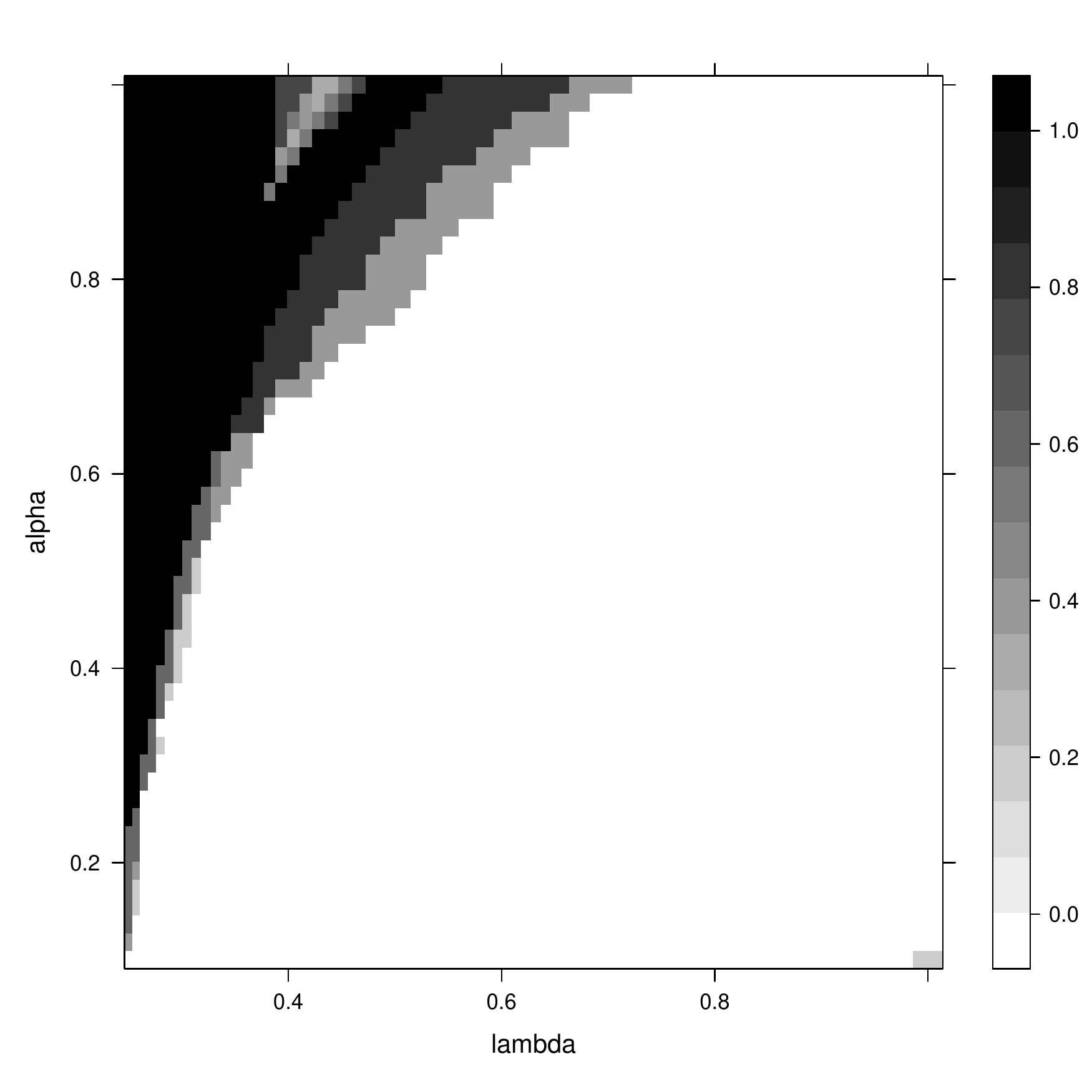,width=1.25 in,height=1.25 in,angle=0} &
			\psfig{figure=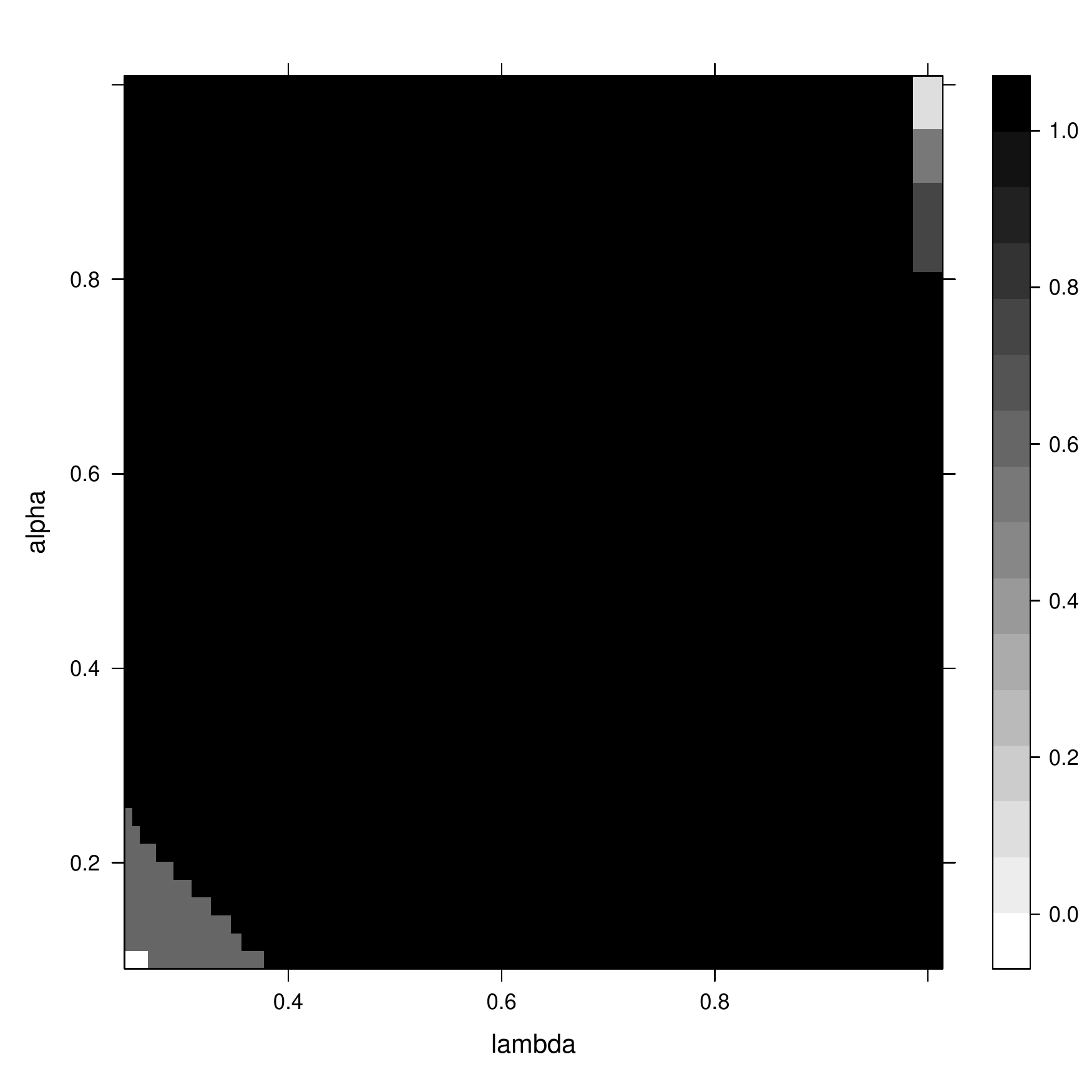,width=1.25 in,height=1.25 in,angle=0} &
			\psfig{figure=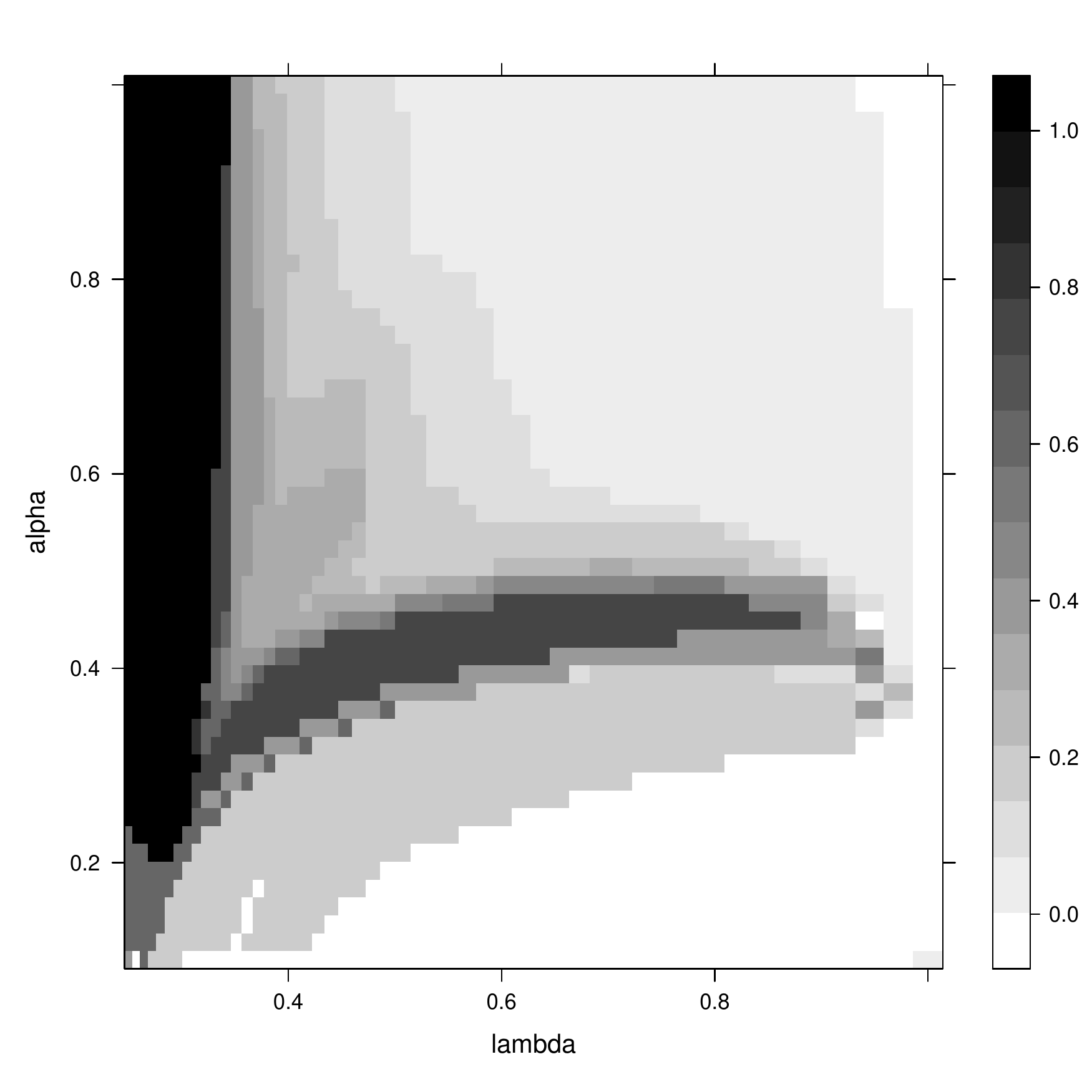,width=1.25 in,height=1.25 in,angle=0} \\
			$\ell_1+\ell_2$ &  $\ell_1+\ell_\infty$ &  $\ell_1+\ell_1/\ell_\infty$&  $\ell_1+\ell_*$\\		
			\hline
		\end{tabular}
	}
	\caption{The critical success index for different models and different penalties with 2500 tuning parameters.}
	\label{fig1}
\end{figure}

Due to the efficiency of HiQR, we can study a high-dimensional case where $p=200$ and $n=500$. It is noted that the model has about { $2\times 10^4$ parameters}. We implement the penalized quadratic regression with 50 $\alpha$s and 50 $\lambda$s, resulting in a solution path for 2500 grids. To measure each estimation $\widehat{\B}$, we adopt the critical success index (CSI), which can evaluate the support recovery rate and the model size simultaneously. For the true $\B$ and an estimation $\widehat{\B}$, the CSI is defined as follows:
\begin{align*}
	\mbox{CSI}(\B,\widehat{\B})=\frac{\#\{(i,j): B_{ij} \neq 0~\mbox{and}~\hat{B}_{ij} \neq 0 \}}{\#\{(i,j): B_{ij} \neq 0~\mbox{or}~\hat{B}_{ij} \neq 0 \}}.
\end{align*}
Figure \ref{fig1} presents the results for different models and different penalties.  From these solution paths, we can see that these methods can detect the true signals if the tuning parameters are set suitably. Tuning parameters selection is beyond the scope of the current work. Our results indicate that the proposed ``HiQR" algorithm is capable of training a model with $2 \times 10^4$ parameters and 2500 tuning parameters efficiently.

\section{Discussions}
In this work, we propose an efficient algorithm for high-dimensional quadratic regression that leverages the special matrix structure of interaction terms. By exploiting the Woodbury identity trick and the properties of the Kronecker product, we derive an explicit solution for ridge-penalized quadratic regression. We then incorporate this solution into the ADMM algorithm to effectively solve the regularized model with non-smooth penalties.
 Building upon the efficient solution for ridge regression, a potential extension of the current work is to address distributed computing scenarios. This would involve adapting the algorithm to handle data distributed across multiple computing nodes. Furthermore, while we employed the classical ADMM algorithm in this study, incorporating computational tricks from the "OSQP" algorithm \citep{osqp} could lead to further enhancements in terms of computational efficiency and scalability. We view these aspects as promising future directions for our research.

 \section*{Acknowledgments}
 We are grateful to the Editor, the Associate Editor and the two referees for their constructive comments, which helped us to improve the manuscript. 
Wang's research is partially supported by NSFC 12031005, NSF of Shanghai 21ZR1432900 and the fundamental research funds for the central universities.  Jiang's research is partially supported by the National Natural Science Foundation of China (12001459), and HKPolyU Internal Grants.

\bibliography{ref}
\end{document}